\documentclass[12pt,preprint]{aastex}

\shorttitle{Suzaku Observation of BD~+30$^{\circ}$~3639}
\shortauthors{Murashima et al.}
\begin{document}
\title{{\it Suzaku} Reveals Helium-burning Products in the\\ 
X-ray Emitting Planetary Nebula BD~+30$^{\circ}$~3639
}

\author{M.~Murashima\altaffilmark{1}, M.~Kokubun\altaffilmark{1},
K.~Makishima\altaffilmark{1,2}, J.~Kotoku\altaffilmark{3},
H.~Murakami\altaffilmark{4}, K.~Matsushita\altaffilmark{5},
K.~Hayashida\altaffilmark{6}, K.~Arnaud\altaffilmark{7},
K.~Hamaguchi\altaffilmark{7}, and H.~Matsumoto\altaffilmark{8}}
\affil{1: Department of Physics, University of Tokyo, 
    7-3-1 Hongo, Bunkyo-ku, Tokyo 113-0011, Japan}
\affil{2:The Institute of Physical and Chemical Research, 
    2-1 Hirosawa, Wako, Saitama 351-0198, Japan}
\affil{3:Department of Physics, Tokyo Institute of Technology,
    O-okayama, Meguro-ku, Tokyo 152-8551, Japan}
\affil{4:PLAIN Center,  ISAS/JAXA,
    3-1-1 Yoshinodai, Sagamihara, Kanagawa 229-8510, Japan}
\affil{5:Department of Physics, Tokyo University of Science,
    Kagurazaka, Shinjuku-ku, Tokyo 162-8601, Japan}
\affil{6:Department of Space and Earth Science, Osaka University,
     Toyonaka, Osaka 560-0043, Japan}
\affil{7:Exploration of the Universe Division, Code 660, 
NASA/GSFC, Greenbelt, MD 20771, USA}
\affil{8:Department of Physics, Kyoto University,
Kitashirakawa, Sakyo-ku, Kyoto 606-8502, Japan}

\begin{abstract}
BD~+30$^{\circ}$~3639, the brightest planetary nebula at X-ray energies,
was observed with {\it Suzaku}, an X-ray observatory launched on 2005 July 10.
Using the X-ray Imaging Spectrometer,
the K-lines from C~{\sc VI}, O~{\sc VII}, and O~{\sc VIII} were  
resolved for the first time,
and C/O, N/O, and Ne/O abundance ratios determined.
The C/O and Ne/O abundance ratios exceed the solar value 
by a factor of at least 30 and 5, respectively.
These results indicate that the X-rays are emitted mainly by
helium shell-burning products.
\end{abstract}


\section{Introduction}

Intermediate-mass stars,
with initial masses $\lesssim 8\ M_{\odot}$,
are thought to contribute significantly to the synthesis of C, N, O, and Ne
through the CNO cycle and He burning.
These nuclear fusion products are ejected through mass loss
as the stars evolve from their AGB (asymptotic giant branch) phase 
into planetary nebulae (PNe).
Hence, a PN can be regarded as a messenger bearing
information on the nucleosynthesis within these stars.
However, the optically-visible material in PNe
represents matter accumulated over the PN lifetime,
making it difficult to extract information on, for instance, 
pure He-burning products just by observing the PNe shells.

Soft X-rays, detected  from several PNe,
are thought to originate in hot plasmas
produced by shocks in fast stellar winds 
\citep{kwok1982,volk1985}.
These fast winds develop during the
later evolutionary stages of the central star,
so the X-rays are thought to be emitted by the star's late-phase products.
X-ray spectroscopy of PNe will thus allow us 
to diagnose products of a particular nucleosynthesis phase 
inside intermediate-mass stars.
However, it has been difficult to resolve K-lines from C, N, and O,
the major CNO and He-burning products, 
because the line energies were too low for X-ray CCDs.
In addition, these objects are  often too X-ray faint for 
high-resolution spectroscopy using gratings,
even when their angular size is small enough.

The 5th Japanese X-ray satellite {\it Suzaku} 
({\it Astro-E2}; Mitsuda et al. 2004)
was developed by a Japan-US collaboration
as a successor to {\it ASCA},
and was launched successfully on 2005 July 10.
Its X-ray Imaging Spectrometer (XIS) 
comprises four CCD cameras, XIS-0 through XIS-3 \citep{XIS}.
Three of the cameras use front-illuminated (FI) CCD chips while
XIS-1 utilizes the back-illuminated (BI) technology.
Compared to similar instruments on preceding X-ray missions,
both the BI and FI chips have better responses
to X-rays with energies below $\sim 1$ keV:
their FWHM energy resolution at 0.5 keV is 
$\sim 40$ eV (FI) and $\sim 50$ eV (BI),
with insignificant low-energy tails.

We observed the PN, BD~+30$^{\circ}$~3639,
with {\it Suzaku} in order to resolve the anticipated 
carbon and oxygen emission lines.
Since its detection using {\it ROSAT} \citep{kreysing1992},
this object has been known as the X-ray brightest PN,
emitting spatially-extended soft X-rays from inside
its $\sim 4''$  diameter optical shell \citep{kastner2000}.
The X-ray emitting plasma is thought 
to have highly non-solar abundance ratios,
as shown by the strong Ne-K line
first detected using {\it ASCA} \citep{arnaud1996},
and a broad sub-keV spectral hump,
presumably a blend of C, N, and O lines,
subsequently detected with {\it Chandra}
\citep{kastner2000,maness2003}.

\section{Observation}
The present  observation of BD~+30$^{\circ}$~3639 
was performed for a net exposure of 34.4 ksec on 2005 September 20,
(\dataset [ADS/Sa.SUZAKU#X/100025010] {seq. no. 100025010}), 
as part of the initial performance verification of {\it Suzaku}.
Both the XIS and the Hard X-ray Detector (HXD; \cite{HXD})
onboard were operated in their nominal modes.
We do not however use the HXD data, 
since its non-imaging $34' \times 34'$  field of view
also contains part of the large supernova remnant,
G65.2+5.7, which overlaps with the target PN.

\begin{figure}
\plotone{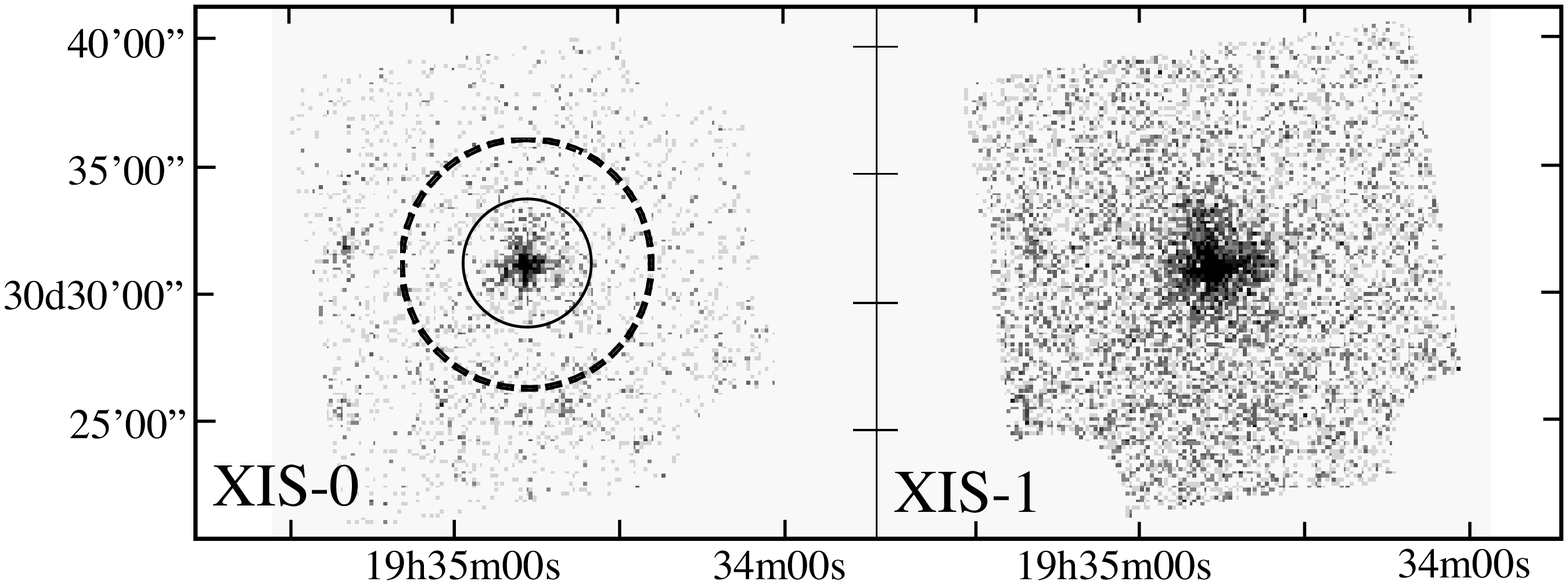}
\caption{Images of BD~+30$^{\circ}$~3639 in 0.3--0.7 keV,
taken with XIS-0 (FI-CCD; left) and XIS-1 (BI-CCD; right).
Two circles specify  on-source and background
data accumulation regions.
}
\label{szimage}
\end{figure}

The target was clearly detected in all XIS cameras;
Figure~\ref{szimage} shows 0.3--0.7 keV images 
obtained with  two of them.
We define the source region as a $2'.5$ radius circle centered on the source,
which encloses  $\sim 90\%$ of signal photons from the PN
that is essentially a point source at the angular resolution of the {\it Suzaku} X-ray Telescopes.
The background region is defined as
 a surrounding annulus with an outer radius of $5'$,
as indicated in Figure~\ref{szimage}a.
After background subtraction, net signal counts are
$1116 \pm 46$, $3039 \pm 74$, $1174 \pm 47$, and $1012 \pm 44$
from XIS-0, 1, 2, and 3, respectively.
Since the three FI cameras (XIS-0, 2, 3) 
are essentially identical, we co-add their data, and refer to them as
``XIS-023'' below.

\section{Analysis and Results}

\begin{figure}
\epsscale{0.9}
\plotone{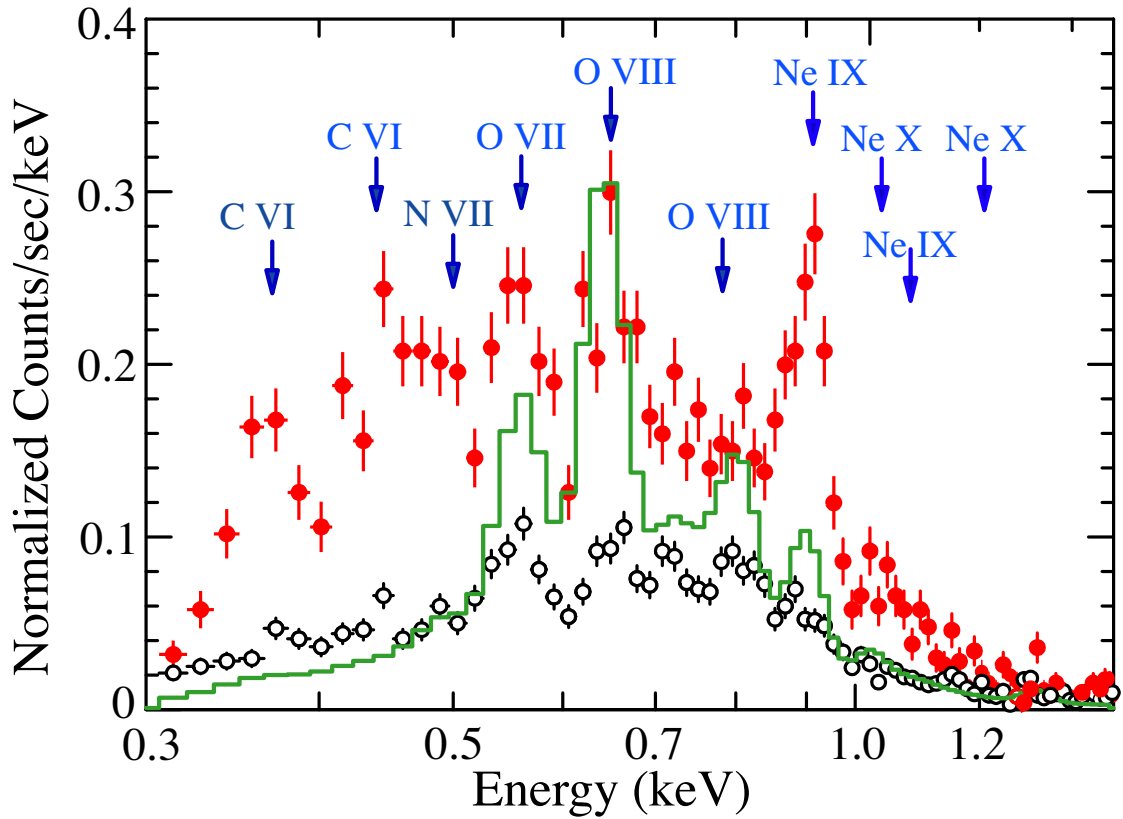}
\caption{Spectra obtained with XIS-1 (BI).
  Red filled circles show the on-source (background inclusive) data,
  and black open circles the background.
  The prediction of an absorbed 1-solar abundance model 
  with $kT=0.2$ keV  is sketched in green.
  Positions of major lines are indicated in blue.
  }
\label{x1spec}
\end{figure}

As shown in Figure \ref{x1spec},
the on-source XIS-1 spectrum clearly reveals at 0.37 keV
the K$_\alpha$ line from hydrogenic carbon (C~{\sc VI}).
Its absence in the background spectrum ensures
that  the line is coming from the PN itself.
The spectra also show K$_\alpha$ lines 
from O~{\sc VII}, O~{\sc VIII}, and Ne~{\sc IX}. 
Some of these lines are found in the background spectrum as well,
but with much reduced intensities;
these background lines are thought to come from  
diffuse Galactic soft X-ray components
which may well extend to the present Galactic latitude of $5.^\circ 0$,
possibly with additional contribution from G65.2+5.7.

At a temperature of $kT \sim 0.3$ keV \citep{arnaud1996,maness2003},
the C~{\sc VI} to O~{\sc VII}   (the triplet summed)
line emissivity ratio of a solar-ratio plasma is $\sim 0.15$.
As illustrated in Figure~\ref{x1spec} in green,
this ratio will be reduced to $\sim 0.03$ \citep{mioPhD}
due to interstellar plus  intra-nebular absorption
(with the latter probably dominant; \cite{kastner2002}),
and the decrease in XIS-1 efficiency toward lower energies.
Nevertheless, the XIS-1 spectrum yields comparable numbers of C~{\sc VI} 
and O~{\sc VII} line photons, 
suggesting a highly enhanced C/O ratio of the X-ray emitting plasma.

This early in the mission, 
the XIS calibration is still being fine-tuned. 
There are uncertainties in the energy gain, 
and also in the low-energy efficiency
which has decreased since the start of observations. 
We solved the former by self-calibrating (within $\sim 3\%$) the gains
using the emission lines from BD~+30$^{\circ}$~3639 themselves.
The latter effect is attributed to excess absorption 
due to the build-up of contaminant in front of the XIS cameras. 
We estimated the chemical composition of the excess absorber
using a {\it Suzaku} observation of the isolated  neutron star RX~J1856.6-3574
\citep{mioPhD},
which is known to exhibit a blackbody with a temperature of 63 eV,
absorbed by a low column of
$N_{\rm H} = 0.087 \times 10^{21}$ cm$^{-2}$ \citep{burwitz2003}. 
To accurately determine the contaminant thickness 
for XIS-1 and (separately) XIS-023, 
we included in all analysis a simultaneous fit to the archival 
(\dataset [ADS/Sa.CXO#obs/00587] {ObsId 587}) 
background-subtracted {\it Chandra} ACIS-S spectrum
of the same object.

\begin{figure}
\epsscale{0.95}
\plotone{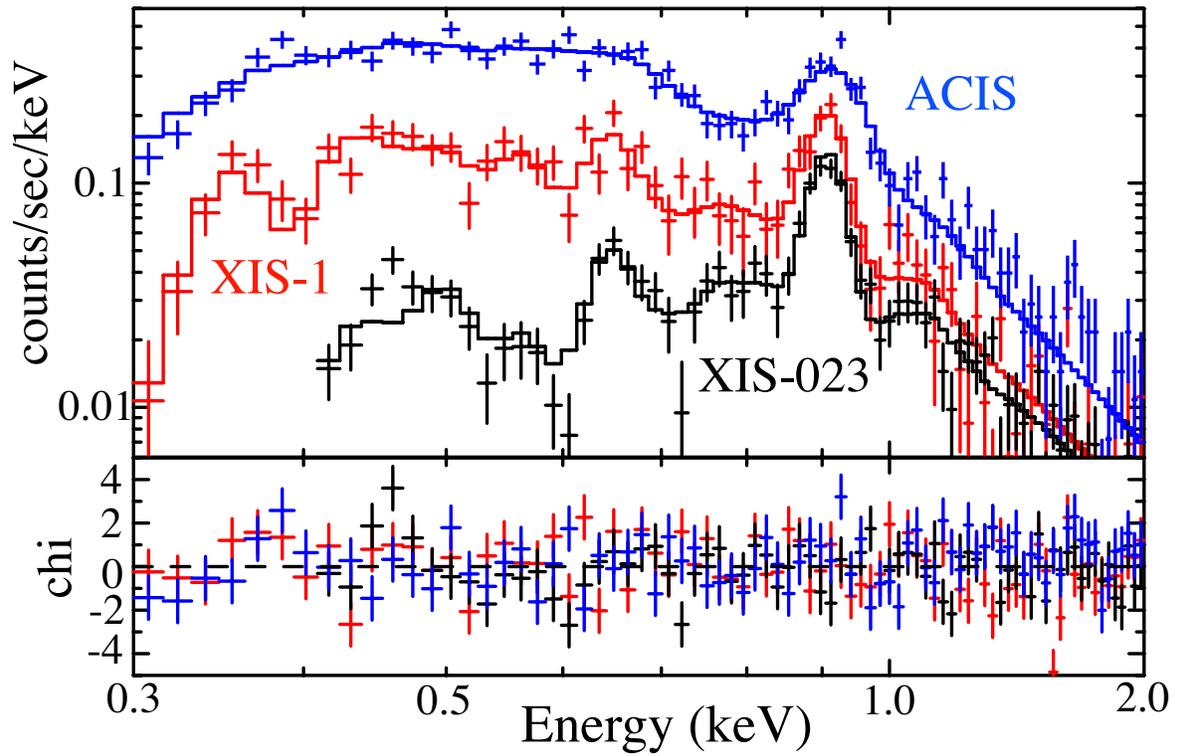}
\caption{Background-subtracted XIS-1 (red), XIS-023 (black), 
 and {\it Chandra} ACIS-S  (blue) spectra of  BD~+30$^{\circ}$~3639,
  fitted jointly by an absorbed vAPEC model.
  The XIS responses take into account the excess absorption.
 }
\label{combined-fit}
\end{figure}

We jointly fit the background-subtracted XIS-1, XIS-023, and 
{\it Chandra} ACIS-S spectra, 
using an absorbed single-temperature vAPEC model \citep{APEC}.
The abundances (relative to solar; \cite{abundances}) of C, N, O, Ne, and Fe,
were left free; that of He fixed to solar; other heavy elements were neglected;
$kT$ and the 
intervening (interstellar plus intra-nebular)
hydrogen column density $N_{\rm H}$  were left free.
As shown in Figure \ref{combined-fit},
the model gives a  reasonable joint fit to the three spectra,
with $\chi^{2}/\nu= 331/228$.
We determined 
$kT=0.19 \pm 0.01$ keV and
$N_{\rm H} = (2.1^{+0.4}_{-0.7}) \times 10^{21}\ {\rm cm}^{-2}$.
Figure~\ref{combined-fit} also illustrates the difference
in the energy-resolving power between the XIS and ACIS-S.

\begin{figure}
\epsscale{0.8}
\plotone{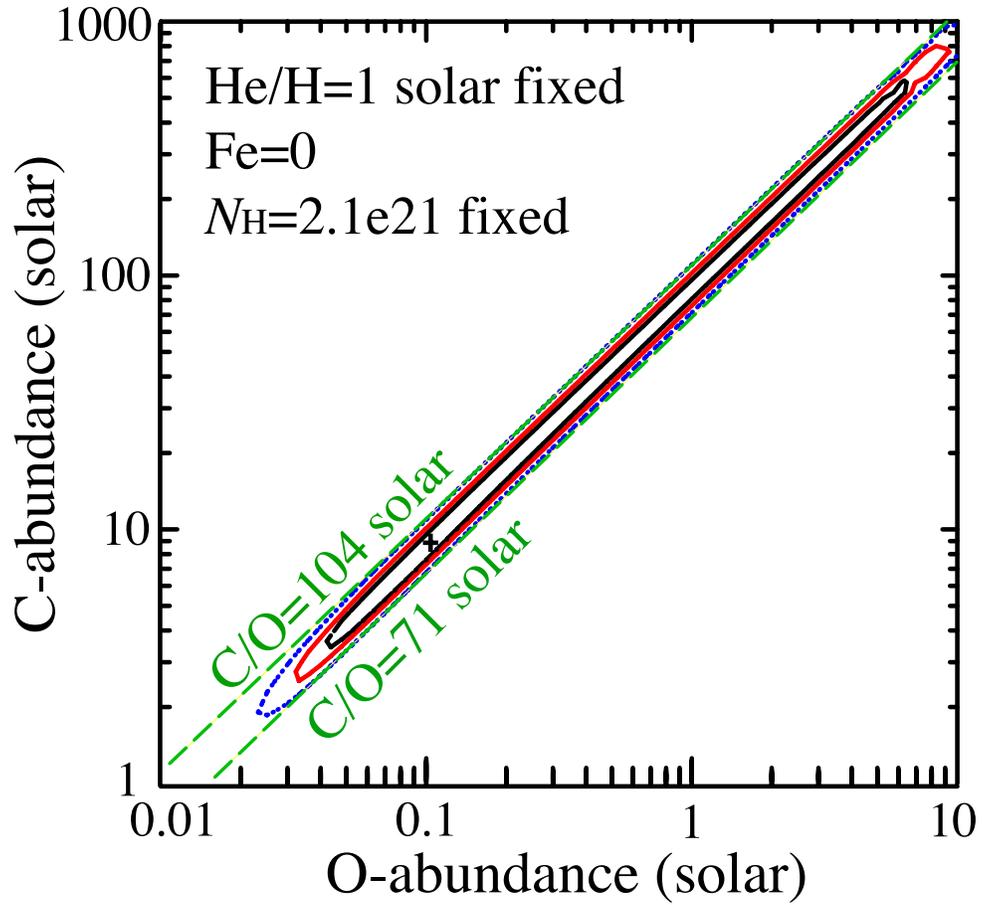}
\caption{Confidence contours for the O and C abundances,
derived from the single vAPEC fit of Figure~\ref{combined-fit}.
The 69\% (black solid), 90\% (red solid), and 99\% (blue dotted)
confidence contours are drawn.
The other parameters are left free,
except for those described in the figure.
}
\label{contour}
\end{figure}

This fitting procedure cannot in reality constrain
the absolute (i.e., relative to hydrogen) metal abundances,
since the low-energy flux from a plasma with
$kT \sim  0.2$ keV is dominated by
emission lines from the metals themselves,
even when the metallicity is $\sim 1$ solar.
Furthermore, the continua, due to hydrogen, 
(presumably non-solar abundance) helium,
and the enhanced metals themselves, 
cannot be independently determined.
Nevertheless, we expect abundance {\it ratios} among  metals
to be  reasonably constrained by the  lines 
which are individually resolved with the XIS.
To see this, we produced in Figure~\ref{contour}
confidence contours for the C and O abundances
using the joint vAPEC fits.
Thus, the  contours are highly elongated
along the  proportionality between the two quantities,
and the C/O ratio is successfully constrained
as $104 >$ C/O $>71$  solar at 90\% confidence
with the best estimate at 85.
(The range does not change significantly even allowing $N_{\rm H}$ to float.)
In the same way, we have obtained $5.5>$N/O
$>0.9$ and  $7.5>$ Ne/O $>4.7$, with the best fits of 3.2 and 5.8, respectively,
and Fe/O$<0.1$, all in solar units.
Below,  these results are examined for various systematic effects.

We  have so far assumed the He/H ratio to be solar.
However, increasing it to 5, 10, or 20 solar
does not affect the relative C/N/O/Ne abundances by more than  5\%.
Similarly, the results remain unchanged within $\sim 10$\% 
when the thickness of the XIS excess absorber
is varied by $\pm 20$\% around the best estimates
(equivalent to $N_{\rm H} \sim 8 \times 10^{20}$ cm$^{-2}$),
or the background spectra are derived from several other regions 
 (e.g., using larger annuli or using limited azimuthal sectors)
in the XIS field of view.

To relax the assumption of isothermality,
we refitted the three spectra jointly with a sum of two vAPEC components,
which are constrained to share common abundances and the same $N_{\rm H}$.
We tentatively fixed the C/O ratio at various trial values,
and let the two temperatures vary freely.
Even with this extra degree of freedom,
acceptable fits were reproduced only
when the assumed C/O ratio is in the range 35--120 solar.
The two temperatures were typically found
at $\sim 0.12$ keV and $\sim 0.22$ keV.
Further, the C/O ratio stayed  above $\sim 30$
even when fitted with a model allowing non-equilibrium 
ionization conditions. 


The largest systematic effect on our results 
comes from uncertainty in $N_{\rm H}$, 
because our best-fit value ($N_{\rm H} \sim 2 \times 10^{21}$ cm$^{-2}$) 
is twice that estimated from extinction data 
at longer wavelengths \citep{cahn1992,arnaud1996,kastner2000}.  
We therefore repeated the joint, single-vAPEC fit to the three spectra,
but this time fixing $N_{\rm H}$ at $1 \times 10^{21}$ cm$^{-2}$.
Then, $kT$ increased to $\sim 0.23$ keV, 
and the C/O ratio decreased by a factor of $\sim 3$,
while the Ne/O ratio remained nearly unchanged. 
These new parameters are consistent with the recent 
{\it Chandra} LETG results on BD~+30$^{\circ}$~3639 \citep{kastner2006}.  
However, the (non-reduced) $\chi^2$ for the joint fit to 
the {\it  Suzaku} XIS and {\it Chandra} ACIS data increases by $\sim 25$
when $N_{\rm H}$ is forced to take the smaller value.  
This suggests that the X-ray absorption is actually higher 
than is implied by the extinction data, 
perhaps because the intra-nebular absorber is metal-enriched.

Based on these evaluations, 
we take as our result the original single-vAPEC solution 
(i.e., C/O $\sim 85$) with free $N_{\rm H}$,
with the reservation that the C/O ratio could be subject to a
systematic over-estimation by a factor of $\sim 3$.  
This factor reflects possible non-isothermality, 
non-equilibrium ionization conditions, 
or the uncertainty in $N_{\rm H}$.  
In any event, the very high carbon over-abundance 
relative to oxygen is a robust result,
confirming the previous suggestions 
\citep{arnaud1996, kastner2000, maness2003}.

\section{Discussion}


The three spectra analyzed consistently indicate an absorbed 
0.2--2.0 keV flux of $6.5 \times 10^{-12}$ erg s$^{-1}$,
in agreement with previous measurements \citep{maness2003}.
Correction for the absorption by $N_{\rm H}=2 \times 10^{21}$ cm$^{-2}$ 
increases this flux by a factor of $\sim 9$,
yielding an absorption-corrected 0.2--2.0 keV luminosity of 
 $\sim 1.4 \times 10^{33}$ ergs s$^{-1}$
at a distance of 1.3 kpc \citep{mellema2004}. 
(If the fit with $N_{\rm H}=1 \times 10^{21}$ cm$^{-2}$ is used,
the absorption-corrected quantities become smaller 
than these by a factor of $\sim 3$.)
This is only $\sim 0.1\%$ of the kinetic luminosity,
$\sim 1.2 \times 10^{36}$ ergs s$^{-1}$,
supplied by the fast stellar wind \citep{mdot}.
This, together with the X-ray emitting region
just fitting inside the optical shell \citep{kastner2000},
supports the interacting wind model \citep{kwok1982,volk1985}
as a scenario to explain the X-ray emission from this PN.

Observations in the optical and neighboring wavelengths
give the nebular abundance ratios of BD~+30$^{\circ}$ 3639
as C/O$\sim 3.7$, N/O$\sim 1.8$,  and Ne/O $\sim 2.8$ 
in solar units \citep{bernard2003}.
Compared with these,
our X-ray results imply qualitatively similar,
but much more extreme, abundance patterns.
This is not surprising,
since the X-ray emitting plasma is considered to
represent a very limited radial zone of the stellar interior.
In fact, assuming, based on the  {\it Chandra} image, 
that the X-ray emitting plasma fills a sphere of radius 0.01 pc ($1''.6$),
its mass is estimated as  $ \sim  5 \times 10^{-4} \: ~ M_{\odot}$
from the observed emission measure of $5.7 \times 10^{55}$ cm$^{-3}$.
This can be supplied only in $\sim 70$ yr
by the mass loss of $\sim 7 \times 10^{-6}~M_{\odot}$ yr$^{-1}$  
from the central star \citep{mdot}.

Which part of the stellar interior provides the X-ray emitting plasma?
The most outstanding  feature is the extreme carbon enhancement over oxygen,
by a factor of $\sim 40$ in the {\it number} ratio, 
or $\sim 30$ in the {\it mass} ratio.
This is a typical value expected from competition 
between the triple-$\alpha$ and $^{12}$C($\alpha, \gamma$)$^{16}$O 
reactions during  He shell-burning flashes \citep{herwig,suda2004},
for an initial mass of $\sim 2~M_{\odot}$.
Therefore, we presume that the X-rays come 
primarily from products of He shell-burning episodes 
which took  place in the AGB  phase.
The possible  factor-of-three  uncertainty
will not affect this inference significantly,
since the theoretical calculations are 
thought to be uncertain to a similar degree.

The observed high Ne/O ratio, in contrast,
requires a neon-producing path without $^{16}$O.
Presumably $^{14}$N, 
concentrated through  the CNO cycle, burnt  as 
$^{14}$N($\alpha, \gamma$)$^{18}$F($\beta ^+, \nu$)$^{18}$O($\alpha, \gamma$)$^{22}$Ne during the He shell flashes.
This produces $^{22}$Ne, 
which is indistinguishable in X-rays from the more abundant  $^{20}$Ne.
Finally, the  hint of enhanced N/O ratio can be explained
if  some fraction of the nitrogen from the CNO cycle,
stored in the upper He layer, 
escaped without being caught up in the He shell burning.

These considerations lead us 
to the following scenario for this PN.
The hydrogen-rich envelope is likely to have already been expelled.
The outer parts of the He layer, 
composed of radiative and convective zones, 
are currently being ejected in the form of fast stellar winds
with a velocity of $\sim 1000$ km s$^{-1}$.
The winds carry the abundant carbon and $^{22}$Ne 
produced in He shell flashes \citep{herwig},
as well as unprocessed nitrogen.
These ejecta are shock-heated to emit X-rays.

A more quantitative comparison with the optical spectra 
of the central stars of PNe \citep{leuenhagen}
will provide a valuable calibration
of the models of stellar evolution and nucleosynthesis.
Such work may also allow us to 
estimate the initial progenitor mass,
and to clarify when the fast stellar winds start blowing. 
In closing, the present {\it Suzaku} observation 
has provided a valuable opportunity of
witnessing the carbon nucleosynthesis  inside an evolved star,
which is one of the central building blocks of the modern astronomy
but has rarely been observed directly.

\acknowledgments

MM, together with the co-authors,
 expresses her deepest gratitude to the
{\it ASTRO-E/Astro-E2/Suzaku}  project members.
Authors  thank Prof. Masayuki~Fujimoto and Dr. Takuma Suda
of Hokkaido University for elucidating discussion.
They are also grateful to Dr. Joel Kastner of Rochester Institute of Technology
for communicating their latest {\it Chandra} LETG results on  BD~+30$^{\circ}$~3639.

{\it Facilities:} \facility{Suzaku (XIS)}.

\clearpage

\clearpage

\clearpage

\clearpage

\end{document}